\documentclass[iop]{emulateapj}

\usepackage{epsfig} 
\usepackage{natbib}
\usepackage{color}
\usepackage{wrapfig}

\def\simgt{\lower.5ex\hbox{$\; \buildrel > \over \sim \;$}}
\def\simlt{\lower.5ex\hbox{$\; \buildrel < \over \sim \;$}}

\def\amin{\ifmmode^{\prime}\else$^{\prime}$\fi}
\def\asec{\ifmmode^{\prime\prime}\else$^{\prime\prime}$\fi}

\def\simgt{\lower.5ex\hbox{$\; \buildrel > \over \sim \;$}}
\def\simlt{\lower.5ex\hbox{$\; \buildrel < \over \sim \;$}}

\newcommand{\goe}{\stackrel{>}{\sim}}

\newcommand\chandra{{\it Chandra}}

\newcommand\xmm{{\it XMM-Newton}}

\newcommand\integral{{\it INTEGRAL}}

\newcommand\swift{{\it Swift\/}}
\newcommand\nustar{\hbox{\it NuSTAR\/}}
\newcommand\fermi{{\it Fermi\/}}

\newcommand\hesssrc{HESS~J1832$-$093}
\newcommand\xmmsrc{XMMU~J183245$-$0921539}
\newcommand\xmmshort{XMMJ183245}

\newcommand\snr{G22.7$-$0.2}
\newcommand\gmc{G22.6$-$0.2}
\newcommand\eflux{erg\,cm$^{-2}$\,s$^{-1}$}

\shorttitle{}
\shortauthors{}

\begin{document}

\title{{\it NuSTAR} hard X-ray observation of the gamma-ray binary candidate HESS~J1832-093}

\author{Kaya~Mori\altaffilmark{1},  E. V. Gotthelf\altaffilmark{1}, Charles J. Hailey\altaffilmark{1}, Ben J. Hord\altaffilmark{1},
Emma de~O\~{n}a~Wilhelmi\altaffilmark{2}, 
Farid~Rahoui\altaffilmark{3,4}, 
John A. Tomsick\altaffilmark{5}, Shuo Zhang\altaffilmark{6}, Jaesub Hong\altaffilmark{4}, Amani M. Garvin\altaffilmark{1}, 
Steven E. Boggs\altaffilmark{5}, Finn E. Christensen\altaffilmark{7}, William W. Craig\altaffilmark{5,8}, 
Fiona A. Harrison\altaffilmark{9},  Daniel Stern\altaffilmark{10}, William W. Zhang\altaffilmark{11}}  

\altaffiltext{1}{Columbia Astrophysics Laboratory, Columbia University, New York, NY 10027, USA}
\altaffiltext{2}{Institute of Space Sciences (CSIC-IEEC), E-08193 Barcelona, Spain}
\altaffiltext{3}{European Southern Observatory, K. Schwarzschild-Strasse 2, D-85748 Garching bei M\"unchen, Germany}
\altaffiltext{4}{Harvard-Smithsonian Center for Astrophysics, Cambridge, MA 02138, USA}
\altaffiltext{5}{Space Sciences Laboratory, University of California, Berkeley, CA 94720, USA}
\altaffiltext{6}{Kavli Institute for Astrophysics and Space Research, Massachusets Institute of Technology, Cambridge, MA 02139, USA}
\altaffiltext{7}{DTU Space - National Space Institute, Technical University of Denmark, Elektrovej 327, 2800 Lyngby, Denmark}
\altaffiltext{8}{Lawrence Livermore National Laboratory, Livermore, CA 94550, USA}
\altaffiltext{9}{Cahill Center for Astronomy and Astrophysics, California Institute of Technology, Pasadena, CA 91125, USA}
\altaffiltext{10}{Jet Propulsion Laboratory, California Institute of Technology, Pasadena, CA 91109, USA}
\altaffiltext{11}{NASA Goddard Space Flight Center, Greenbelt, MD 20771, USA}

\begin{abstract}

  We present a hard X-ray observation of the TeV gamma-ray binary
  candidate \hesssrc\ coincident with supernova remnant (SNR) \snr\
  using the \textit{Nuclear Spectroscopic Telescope Array}
  (\textit{NuSTAR}).  Non-thermal X-ray emission from \xmmsrc, the
  X-ray source associated with \hesssrc, is detected up to
  $\sim30$~keV and is well-described by an absorbed power-law model
  with the best-fit photon index $\Gamma = 1.5\pm0.1$. A re-analysis
  of archival \chandra\ and \xmm\ data finds that the long-term X-ray
  flux increase of \xmmsrc\ is $50_{-20}^{+40}$\% (90\% C.L.), much less than previously reported. A search for a pulsar spin period or binary
  orbit modulation yields no significant signal to a pulse fraction
  limit of $f_p<19$\% in the range 4~ms~$< P<40$~ks.  No red noise is
  detected in the FFT power spectrum to suggest active accretion from
  a binary system.  While further evidence is required, we argue that
  the X-ray and gamma-ray properties of \xmmsrc\ are most consistent with a
  non-accreting binary generating synchrotron X-rays from particle
  acceleration in the shock formed as a result of the pulsar and
  stellar wind collision.  We also report on three nearby hard X-ray
  sources, one of which may be associated with diffuse emission from a
  fast-moving supernova fragment interacting with a dense molecular
  cloud.

\end{abstract}
\keywords{}


\section{Introduction}
\label{sec:intro}

High energy gamma-ray surveys using ground-based Cherenkov telescopes
(e.g., MAGIC, H.E.S.S, and VERITAS) have uncovered a rare subclass of
TeV binary systems \citep{Dubus2013, Dubus2015b}. Whereas the majority
of the $\sim80$ identified Galactic TeV sources are associated with
either pulsar wind nebulae (PWNe) or supernova remnants (SNRs), six
gamma-ray binaries have been detected above $E\sim100$~GeV. These
include, PSR~B1259-63, LS~5039, LS~I$+$61~303, HESS~J0632$+$057,
1FGL~J1018.6$-$5856 and, most recently, PSR~J2032$+$4127
\citep{Aharonian2005, Aharonian2006, Albert2009, Aharonian2007,
  Hess2015, Lyne2015}.  These sources are all identified as
non-accreting binaries harboring an O or B main sequence star and a
compact object, with a wide range of orbital periods from 3.9~days to
$\sim50$~years.  With the exception of PSR~B1259$-$63 and
PSR~J2032$+$4127, both of which are known to have radio pulsars, the
nature of the compact object, whether a neutron star (NS) or black
hole (BH), remains unknown.

\begin{figure*}
\begin{center}
\mbox{
\epsfig{figure=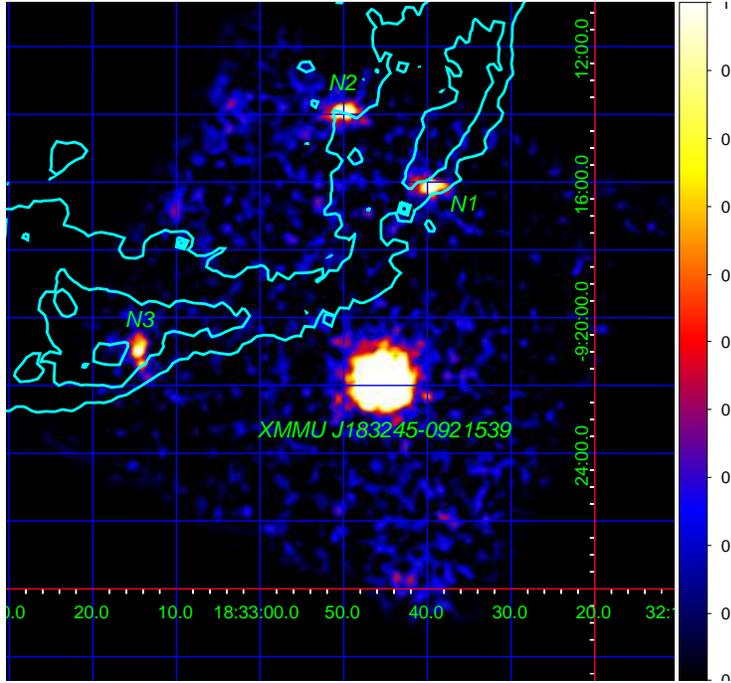,height=0.5\linewidth}
}
\caption{Background-subtracted \nustar\ 3-30~keV image overlaid with 20~cm radio (cyan) contours
of the
SNR~shell G22.7$-$0.2 \citep{Helfand2006}. We
combined module A and B images after subtracting background models generated by {\tt nuskybgd}.
The image was smoothed by a Gaussian kernel with a 5-pixel (12.5\asec) width. The image shows
the X-ray counterpart of \hesssrc\  and three other X-ray sources (N1, N2 and N3).  }
\label{fig:nustar_image}
\end{center}
\end{figure*}

\begin{figure*}
\begin{center}
\mbox{
\epsfig{figure=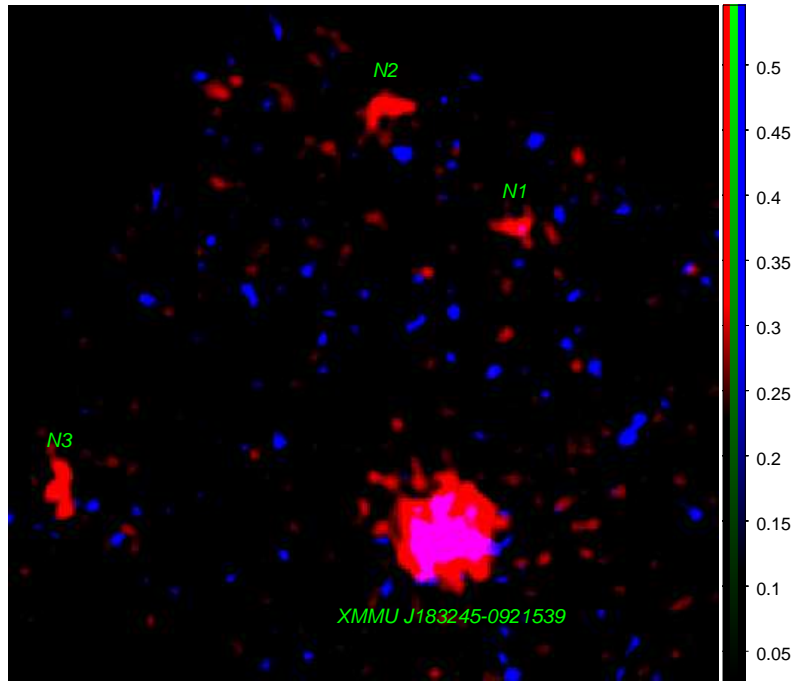,height=0.5\linewidth}
}
\caption{Two-color \nustar\ images in 10-20 (red) and 20-30 (blue) keV bands. We
combined module A and B images after subtracting background models generated by {\tt nuskybgd} and smoothing by a Gaussian kernel with a 5-pixel (12.5\asec) width. The image was zoomed in 
the X-ray counterpart of \hesssrc\  and three other X-ray sources (N1, N2 and N3).  }
\label{fig:nustar_image_10-30keV}
\end{center}
\end{figure*}

Gamma-ray emission from TeV binaries are generally thought to
originate from particle acceleration in the shock formed between the
stellar wind and the pulsar wind \citep{Tavani1994, Sierpowska2008,
  Dubus2013}. Both X-ray and gamma-rays result from synchrotron and
inverse Compton scattering in the interaction region, respectively,
and show a strong dependence on orbital phase.  Other scenarios such
as the microquasar model \citep{Romero2003, Bosch-Ramon2004} are less
plausible since the spectral and timing properties are similar in all
TeV gamma-ray binaries, including the two containing radio pulsars
\citep{Dubus2013}.  Multi-wavelength monitoring of the gamma-ray
binaries in the X-ray, GeV and TeV bands revealed the complex emission
mechanisms and geometry \citep{Kaspi1995, Chernyakova2006b,
  Chernyakova2009, Uchiyama2009, Chernyakova2006, Takahashi2009,
  Kishishita2009, Rea2011b, Aliu2014, Li2011, An2013, An2015, Ho2017}.
A number of theoretical models, including numerical hydrodynamics
simulations, have been developed to account for the orbital dependence
of the X-ray and gamma-ray spectra via anisotropic radiation
processes, relativistic Doppler boosting and inhomogeneous stellar
winds \citep{Dubus2015a, Takata2017, Cita2017}.  Studying these rare
gamma-ray binaries not only probes the unique environment of the
pulsar and stellar winds interaction but also sheds light on the
(short) X-ray binary evolution stage before they become ``regular''
accretion powered high-mass X-ray binaries (HMXBs) \citep{Dubus2013}.

The unresolved TeV point source \hesssrc\ was
discovered in the vicinity of SNR \snr\ suggesting a 
possible association \citep{Abramowski2015}. 
However, follow-up X-ray observations favored a binary origin for the
TeV emission \citep{Eger2016}. The bright X-ray source \xmmsrc\ (\xmmshort\ herein) lies
within the gamma-ray error circle \citep{Abramowski2015} and is
associated with a \chandra\ point source \citep{Eger2016}.  The latter
authors reported a large, factor of 4, increase in the \chandra\ flux
relative to the earlier \xmm\ measurement that seemed to rule out the
PWN or SNR scenario for the X-rays. Instead, the
coincidence of a bright IR source at the
\chandra\ coordinates, a plausible counterpart, suggests  a binary
scenario for powering the gamma-ray emission  \citep{Eger2016}.

In this paper, we present a \nustar\ X-ray observation of the field
containing \hesssrc, along with a re-analysis of the archival \xmm\ and
\chandra\ data. Observational details of these data sets are given in
\S\ref{sec:obs}.  Spectroscopy and timing results on \xmmshort\ are reported
in \S\ref{sec:spec} and \S\ref{sec:timing}, respectively.  Our
analysis of archival \chandra\ data refute reports in previous work of
large flux variations.  Nevertheless, we find sufficient evidence to
prefer the TeV gamma-ray binary scenario for \hesssrc. A timing
analysis detailed in \S\ref{sec:timing} places upper limits on any probable  
pulsar or binary signal. We also present the analysis, in
\S\ref{sec:3sources}, of three nearby hard X-ray sources. We discuss
the nature of \hesssrc\ and the hard sources in
\S\ref{sec:discussion}. Finally, we summarize our results in
\S\ref{sec:summary}.


\section{\nustar\ observation and data reduction}
\label{sec:obs}

An 87~ks {\it NuSTAR} observation of the field containing \hesssrc\
was obtained on 2016 March 21 as part of the \nustar\ TeV survey
project. \nustar\ consists of two co-aligned X-ray telescopes, with
corresponding focal plane modules FPMA and FPMB that provide
$18^{\prime\prime}$ FWHM imaging resolution over a 3--79~keV X-ray
band, with a characteristic spectral resolution of 400 eV FWHM at 10
keV \citep{Harrison2013}.  The reconstructed \nustar\ coordinates are
accurate to $7\farcs5$ at the 90\% confidence level.  The nominal
timing accuracy of \nustar\ is $\sim$2~ms rms, after correcting for
drift of the on-board clock, with the absolute timescale shown to be
better than $<3$~ms \citep{Mori2014, Madsen2015}.

The data was processed and analysed using the {\tt FTOOLS}
09May2016\_V6.19 software package including ({\tt NUSTARDAS}
14Apr16\_V1.6.0) with \nustar\ Calibration Database (CALDB) files of
2016 July 6. No flares were evident during the observation, resulting
a total of 86.9~ks of net usable exposure times spanning 171.5~ks.
For all following spectral analysis, extracted spectra, grouped into
appropriate channels, were fitted using the {\tt XSPEC} (v12.8.2)
package \citep{Arnaud1996}. These fits make use of the {\tt TBabs}
absorption model in {\tt XSPEC} with the {\tt wilm} Solar abundances
\citep{Wilms2000} and the {\tt vern} photoionization cross-section
\citep{Verner1996}. $\chi^2$ statistics were used to evaluate the
spectral fits. All errors quoted herein are for 90\% confidence level
(C.L.).

The \nustar\ background contamination is highly variable across the
focal plane of the two FPM detectors. We use the {\tt nuskybgd}
software \citep{Wik2014} to help model the spatial and energy dependent
cosmic X-ray and a detector background.  This allows us to generate,
for each detector, a model energy-resolved background map for image
analysis and background spectra at the source location for our spectral
analysis. The background model components are normalized by
simultaneously fitting \nustar\ spectra in three source-free regions.
The {\tt nuskybgd} model fit to the source-free spectra yields $\chi^2_\nu = 1.1$ (1472 dof) 
without apparent Fe lines at E $\sim$ 6-7 keV which is indicative of the Galactic ridge X-ray emission 
\citep{Mori2015}. 
In addition, as shown in \S\ref{sec:spec}, \nustar\ module A and B spectra of the brightest 
X-ray source in the FOV (\xmmsrc) jointly fit by an absorbed power-law model show their  
relative flux normalization is 2\%. The 2\% flux discrepancy between the two module spectra 
is not only below the statistical errors ($\sim$3\%) but also it indicates that any additional 
background component unaccounted for by the {\tt nuskybgd} model has a negligible contribution of 
less than 2\% in the 3-30 keV band where all our imaging and spectral analysis are performed.      

Figure~\ref{fig:nustar_image} presents the combining
background-subtracted, exposure-corrected smoothed \nustar\ images from
the two detector modules, in the 3$-$30~keV energy band.  Using {\tt
  wavdetect}, we detected four $>3\sigma$ sources, including the X-ray
counterpart \xmmshort\ to \hesssrc.  Interestingly, the other three
\nustar\ sources (N1, N2 and N3 hereafter), all of which have \xmm\
counterparts, overlap with the radio shell of SNR G22.7$-$0.2 (see
Figure~\ref{fig:nustar_image}, cyan contours).  Above 20~keV, only
\xmmshort\ is visible in the \nustar\ images (see Figure~\ref{fig:nustar_image_10-30keV} for two-color \nustar\ images in 
10-20 and 20-30 keV bands). Based on the 3XMM source
catalog \citep{Rosen2016}, we found that the four \nustar\ sources are
the brightest among about a dozen \xmm\ point sources in the \nustar\
field of view.

We also analyzed archival \xmm\ and \chandra\ observations of the
\hesssrc\ field. A 17~ks \xmm\ exposure (ObsID \#0654480101) was
obtained on 2011 March 13 and an 18~ks \chandra\ pointing (ObsID \#16737)
was acquired on 2015 July 6. Details of these observations
and their analysis can be found in \cite{Abramowski2015} and
\cite{Eger2016}, respectively.  Although \swift\ observations overlap
with the vicinity, their short exposures result in few photons (15-26 cts)
to measure a flux accurately. As reported in \citet{Eger2016}, these
data suggest no evidence of large flux variability among 4 data sets
spanning 2008 Feb 28 to 2015 Sep 26.

{
\tiny 
\begin{deluxetable*}{lcccccc}
\tablecaption{Spectral Results for \xmmsrc}
\tablewidth{0pt}
\tablecolumns{7}
\tablehead{
            \colhead{Data Set} & \colhead{$N_{\rm H}$}          & \colhead{$\Gamma$} & \colhead{Flux\tablenotemark{a}} & \colhead{Flux\tablenotemark{a}} & \colhead{Flux\tablenotemark{a}}& $\chi_\nu^2$ (dof) \\
           \colhead{(Fitted Band, Observation date)} & \colhead{($10^{22}$\,cm$^{-2}$)} & \colhead{}         & \colhead{\xmm}                    & \colhead{\chandra}            & \colhead{\nustar}           & \\
           \colhead{             } & \colhead{}                       & \colhead{}         & \colhead{(EPIC pn)}                & \colhead{(ACIS)}                & \colhead{(FPMA)}            & \\
           \colhead{}              & \colhead{}                       & \colhead{}         & \colhead{(EPIC MOS)}               & \colhead{}                      & \colhead{(FPMB)}            & 
}
\startdata
\xmm\      (2$-$8 keV, 2011 March 13)                       & $9.5$\tablenotemark{b} & $1.0\pm0.3$  & 6.6(5.6-7.2) &  \dots  & \dots & 0.97(41)\\ 
                                             &                         &              & 6.4(5.4-7.1) &  \dots  & \dots & \\ 
\chandra\  (2$-$8 keV, 2015 July 6)                       & $9.5$\tablenotemark{b} & $1.7\pm0.5$  &  \dots       &  7.7(6.1-8.5) &\dots  & 0.81(16)\\ 

\nustar\   (3$-$30 keV, 2016 March 21)                      & $9.5$\tablenotemark{b} & $1.5\pm0.1$  &  \dots & \dots              & 9.6(9.1-10.0) & 1.0(151)\\ 
                                             &                         &              &                     &       & 9.7(9.3-10.1)& \\ 
\chandra\ + \xmm\ (2$-$8 keV)\tablenotemark{c}& $9.7\pm5$               & $1.2\pm0.7$  & 6.2(2.6-6.7) & 8.9(3.8-9.5) & \dots & 1.0(57)\\ 
                                             &                         &              & 5.9(2.6-6.4) &              &       &\\ 
\xmm\ + \chandra\ + \nustar\ (2$-$30 keV)\tablenotemark{c}& $9.5\pm2$  & $1.5\pm0.1$  & $5.8\pm0.5$ & $8.0\pm0.8$ & 9.4(9.0-9.8) & 1.0(112)  \\
                                                         &             &              & $5.2\pm0.6$ &             & 9.6(9.1-10.0) &   
\enddata
\tablenotetext{a}{Absorbed flux for the 2$-$10~keV band in units
  of $10^{-13}$ \eflux. Uncertainties are estimated using the {\tt XSPEC} {\it flux} command for the 90\% confidence level}
\tablenotetext{b}{Column density is fixed to the best-fit value obtained from a simultanous spectral fit to the combined  \xmm\ + \chandra\ + \nustar\ data. Errors are given for 2 interesting parameters at the 90\% C.L. }
\tablenotetext{c}{We linked column density and photon index between the different spectra.}
\label{tab:spectral_fit}
\end{deluxetable*}
}

\begin{figure}[t]
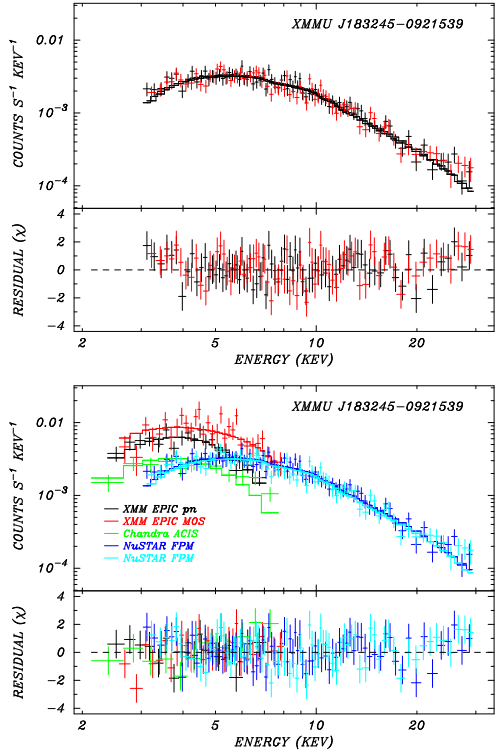

\centerline{
\epsfig{figure=hess1832_nustar_spec_v2.ps,height=0.75\linewidth,angle=270}          
}
\vspace{2mm}
\centerline{
\epsfig{figure=hess1832_all_spec_v2.ps,height=0.75\linewidth,angle=270}          
}
\caption{ {\it Top ---} The \nustar\ 3-30~keV X-ray spectrum of
  \xmmshort\ fitted with an absorbed power-law model. The best-fit model
  (histogram) and data points (crosses) are shown in the top
  panel. Residuals from the best-fit model are shown in the lower
  panel.  {\it Bottom ---} Simultaneous fit to \chandra, \xmm, and
  \nustar\ spectra of \xmmshort, with the column density and power-law
  index parameters linked. The fitted model is given in Table~\ref{tab:spectral_fit}.}
\label{fig:hess_spectral_fit}
\end{figure}

\section{Spectral analysis of \xmmsrc}
\label{sec:spec}

We extracted \nustar\ spectra of \xmmshort\ from a $r=50$\asec\ region
and generated \nustar\ response matrix and ancillary response files
using \textbf{nuproducts}.  We created a model background spectrum for
each of the source regions using the {\tt nuskybgd} tool. As a result,
the flux normalization for spectra extracted from the two modules
match within 2\% of each other. These spectra are grouped with at
least 30 counts in each fitted channel bin.

The {\it NuSTAR} spectra of \xmmshort\ extend up to 30~keV, above
which the background dominates, and is well-fit to an absorbed
power-law model. However, the column density  derived from these
data is found to be unconstrained, with 100\% uncertainties. This is
also the case for individual fits to the \chandra\ and \nustar\ data
on \xmmshort.  Instead, for all subsequent spectral fits to individual
data sets, we hold the column density fixed to $N_{\rm H} =
9.5\times10^{22}$~cm$^{-2}$. This value results from a simultaneous
fit to the \nustar, \xmm, and \chandra\ spectra, as described in the
next section.  For this nominal column density we obtain a best-fit
photon index of $\Gamma = 1.5 \pm 0.1$ for the \nustar\ spectra, with a fit
statistic of $\chi^2_\nu$=1.0 (151 dof). This yields an absorbed
2$-$10~keV flux of $F = 9.6\pm0.8 \times 10^{-13}$ \eflux\ for the
FPMA spectrum and similar result for other module.  No significant spectral break
or cutoff was detected.  A summary of all the spectral results for
\xmmshort\ obtained herein is given in Table~\ref{tab:spectral_fit}.

To explore flux and spectral variations of \xmmshort\ on short
timescales ($\sim 6$~hour), we repeated our spectral fits to \nustar\
data extracted from four equally-divided intervals of the light curve
(20~ks each) in the 3$-$30~keV energy range. However, no significant
change is found in the flux or photon index during the observation. A
similar result is obtained for spectra in the 10$-$30~keV bands, where
any effects of low energy absorption is expected to be negligible. We
conclude that there was no spectral variation during the \nustar\
observation to the limit of measurement uncertainties.

\begin{figure}[t]
\begin{center}
\epsfig{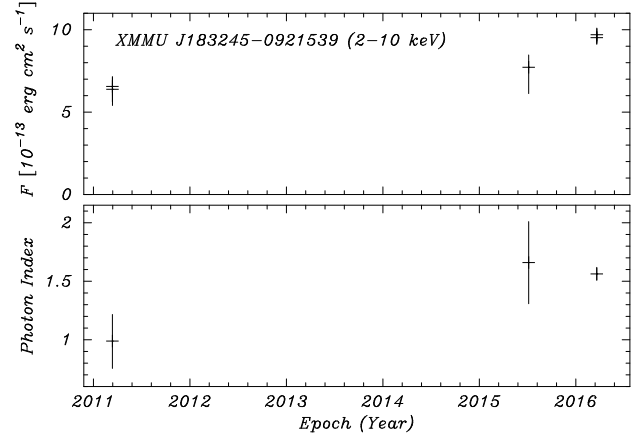}
\caption{The best-fit 2-10 keV absorbed fluxes (top) and X-ray power-law photon indices (bottom) for \xmmshort\
obtained from the \xmm, \chandra\ and \nustar\ observations. The quoted errors are for the 90\% Confidence Level.  Data points are from Table~\ref{tab:spectral_fit}.}
\label{fig:spec_compare}
\end{center}
\end{figure}

To quantify the long-term flux and spectral variability of \xmmshort\ we
compare result of our \nustar\ observations with the archival \xmm\
and \chandra\ data sets spanning a total of 5 years. For each mission,
we follow the standard reduction and analysis procedures. For
\chandra, we used the {\tt specextract} script to extract ACIS source
counts from a $r<3\arcsec5$ region file and to generate the spectrum
and its response files for the point source.  We fit the resulting
spectrum to the absorbed power-law model with the column density fixed
to the nominal value and obtain best-fit photon index of $\Gamma = 1.0\pm
0.3$ and a 2$-$10~keV flux of $F = 6.6(5.6-7.2) \times 10^{-13}$
\eflux.  The magnitude of this flux falls within $\sim2\sigma$ of the
value obtained a year later using \nustar\ data (see
Table~\ref{tab:spectral_fit}) but notably $\sim4$ times less than that
reported in \citet{Eger2016}\footnote{We note that we obtain a similar
  number of extracted counts (409 cts, $0.3<E<10$~keV) for our
  spectrum as recorded by \citet{Eger2016} in their Table~1 (416 cts)}.
As a check, we entered the derived counts rate and spectral parameters
into PIMMS\footnote{\url{http://cxc.harvard.edu/toolkit/pimms.jsp}},
allowing for the 4.9\% deadtime in the ACIS 1/8 subarray observing
mode. This verified the flux result presented here.

For the \xmm\ analysis, we extracted EPIC pn and MOS spectra using a
$r=30$\asec\ aperture around \xmmshort. 
Background spectra were extracted from an annulus region around the
source. A joint fit to the PN and merged MOS spectra in the 2$-$8~keV
band with the nominal absorbed power-law model yielded a photon index of $\Gamma =1.0\pm0.3$
and a 2$-$10 keV flux $F = 6.6(5.6-7.2)\times 10^{-13}$ \eflux\ for
the EPIC pn  spectra and similar for the MOS fits, with a fit statistic of
$\chi^2_{\nu} = 0.97$ for 41~DoF, confirming the results of
\citet{Abramowski2015}.

We use the 2016 \nustar\ flux measurements, along with the corrected
2015 \chandra\ results and the 2011 \xmm\ data to quantify the source
variability of \xmmshort.  Figure~\ref{fig:spec_compare} summarized its
flux and spectral history. The most extreme change is between the
initial and lastest data sets, representing a $\sim 50\%$ fractional
increase in flux, significant at the $5\sigma$ level.  The power-law
photon indices are found to be consistent between observations to
within the measured errors.


\section{Timing analysis of \xmmsrc} 
\label{sec:timing}

We searched for temporal evidence of a binary orbit for \xmmshort\ using
the \xmm, \chandra, and \nustar\ data sets.  Photon arrival times obtained
from each mission were first corrected to the Solar System barycenter
using the JPL DE200 planetary ephemeris and the \chandra\ derived
coordinates of R.A. 18:32:45.158, Decl. $-$09:21:54.78 (J2000).
Initial analysis shows no signature characteristic of an accreting
system; the light curves are stable on all timescales and their fast
Fourier transforms (FFTs) power spectra shows no evidence of red
noise. 
To search for a coherent pulsed signal we used both the FFT
method and the unbinned $Z^2_n$ test, for $n= 1,2,3,5$, and the
H-test, to be sensitive to both broad and narrow pulse profiles.

From the \chandra\ data we extracted $N =399$ counts in a $r
<1.8^{\prime\prime}$ radius aperture, containing essentially no
background contamination ($\leq 1$~count). We performed a Nyquist
limited FFT search in the 0.3$-$10 keV energy band and find no
significant signal for periods between 1.68~s and 10.4~ks, with a
$3\sigma$ upper limit on a sinusoidal pulse fraction of
$f_p(3\sigma)<40\%$ for $2^{15}$ trials. For the \xmm\ observation we
obtained 576~cts and 830~cts from the merged MOS and the pn data sets,
respectively, extracted using a $r<0.4^{\prime}$ source aperture in
the 0.3$-$10~keV energy range. A $2^{16}$ element FFT search of the
MOS data yields no significant signal, with upper limit of
$f_p(3\sigma) < 46.5\%$ for $0.6 < P < 8,310$~s. Similarly, from the
pn data, we obtain an $f_p(3\sigma) < 36.9\%$ between $P=146.8$~ms and
$P=7.5$~ks using a $2^{18}$ element FFT. These upper limits take into
account the estimated background contamination in the source aperture.
We find no evidence for a binary orbit signature in any of these
searches.

The high time resolution \nustar\ data allows a search for coherent
pulsations, as suggested by the X-ray spectrum of \xmmshort, typical
of a young, rapidly rotating pulsar, possibly association with a HESS
source.  From the merged FPM data we extracted $N = 6030$~cts
contained within a $r <0.8^{\prime}$ radius source aperture in the
full \nustar\ energy band. For a $2^{27}$ element FFT we obtain an
upper limit on the pulse fraction to $f_p(3\sigma) < 19.4\%$ between
$P = 4$~ms and $P =85.7$~ks, after allowing for the source aperture
background. We also searched for a signal over a restricted energy
range of $<20$~keV and 20$-$79~keV, however, none was detected.  We
repeated all our searches using the $Z^2_n$ method and H-test which
produce consistent results.

For the long time span (171~ks) of the \nustar\ observation we
performed an additional test using an accelerated FFT search to sample
a range of frequency derivatives typical of a energetic pulsar. In no
case did we detect a significant pulsar or orbital signal.  On the
other hand, unlike for the \chandra\ and \xmm\ data, the strong
signature in the power spectrum at the 97~min \nustar\ spacecraft
orbital period and its many harmonics can mask an adjacent binary
signal in the frequency domain.  The non-detection of X-ray pulsation
is common for gamma-ray binaries with upper limits on the pulsed
fraction from $\sim8$\% to 35\% since the unpulsed wind emission may
be dominant \citep{Hirayama1999, Martocchia2005, Rea2010, Rea2011a,
  Rea2011b}.


\section{Spectral analysis of N1, N2, N3} 
\label{sec:3sources}

To determine the possible nature of the three X-ray sources detected
in the hard band, N1, N2 and N3, we extracted \nustar\ and \xmm\
spectra, using a $r=30$\asec\ aperture.  The \chandra\ observation of
\xmmshort\ was operated in the 1/8 sub-array mode and did not overlap
any of these sources in its restricted field-of-view. Due to the lack
of sufficient counts in the MOS spectra for spectroscopy, after
background subtraction, we only fit the EPIC~pn data. The \nustar\
FPMA data for N2 is heavily contaminated by additional stray-light
background photons from a nearby bright source, and is excluded from
the analysis. We again generated \nustar\ background spectra for each
source using the {\tt nuskybgd} model.  The previous \xmm\ background
spectra proved adequate for their spectral analysis.  After rebinning
the spectra with at least 20 counts per bin, spectral fitting was
performed from $\sim$3~keV to 20~keV where the background is not
significant.

\begin{figure}[t]
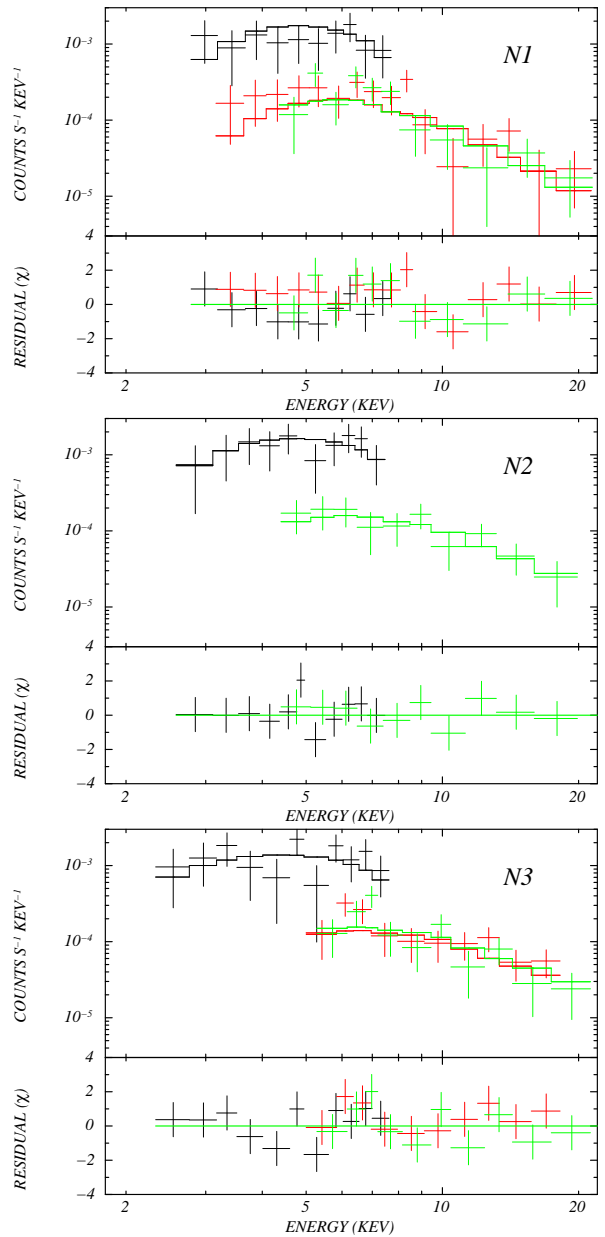

\begin{center}
\epsfig{figure=n1_xmm_nustar_spectra.eps,height=0.95\linewidth, angle=270}

\epsfig{figure=n2_xmm_nustar_spectra.eps,height=0.95\linewidth, angle=270}

\epsfig{figure=n3_xmm_nustar_spectra.eps,height=0.95\linewidth, angle=270}
\caption{\xmm\ + \nustar\ spectra of the three hard X-ray sources (EPIC-PN: black, module A: red, module B: green)
jointly fit by an absorbed power-law model.}
\label{fig:spectral_fit}
\end{center}
\end{figure}

Given that we find no significant flux deviation between the \xmm\ and
\nustar\ observations for the three hard X-ray sources, we jointly fit
the \xmm\ and \nustar\ spectra for each. The fit results obtained for
an absorbed power-law model are summarized in
Table~\ref{tab:spectral_fit_3sources} and
Figure~\ref{fig:spectral_fit}. In addition, we fit an absorbed, 
optically-thin thermal plasma model ({\tt tbabs*apec}). We fixed the
abundance to solar since it is poorly constrained due to the absence
of apparent Fe lines.

A power-law model fit to the \nustar\ + \xmm\ spectra of N1 yields
$N_{\rm H} = 17_{-7}^{+10}\times10^{22}$~cm$^{-2}$ and a photon index $\Gamma = 2.1_{-0.4}^{+0.5}$.
An absorbed optically-thin thermal plasma model ({\tt tbabs*apec}) fits the spectra equally well with  $\chi^2_{\nu} = 1.0$ with
a best-fit
column density and temperature of $N_{\rm H} = (17_{-7}^{+9})\times10^{22}$~cm$^{-2}$ and
$kT = 13_{-5}^{+16}$~[keV]. 
In contrast, \nustar\ + \xmm\ spectra of N2 and N3 fit to a power-law model with harder power-law photon
indices $\Gamma = 1.2\pm0.4$ and $\Gamma = 0.9_{-0.3}^{+0.4}$, respectively. An absorbed APEC model fit yields $kT = 46_{-26}^{+18}$ keV ($\chi_\nu^2=0.57$) and $27_{\
-12}^{+37}$~keV ($\chi_\nu^2 = 1.00$) for N2 and N3, respectively \footnote{The upper bound of plasma temperature is set by the maximum value ($kT = 64$~keV) allowed \
in the APEC model.}. We also added a partial-covering absorption model ({\tt pcfabs}) in order to account for X-ray reflection from the white dwarf surface or absorption in the accretion curtain
for intermediate polars \citep{Hailey2016}. A {\tt tbabs*pcfabs*apec} model fit did not constrain the parameters well
for N2.
The same model fit to the \nustar\ + \xmm\ spectra of N3 ($\chi_\nu^2 = 0.66$) yields a lower temperature $kT = 13_{-3}^{+5}$~keV, partial covering column density $N_H\
 ({\rm pc}) = 10_{-3}^{+5}\times10^{23}$~cm$^{-2}$ and covering factor $f_c = 0.88_{-0.08}^{+0.05}$.

\begin{deluxetable*}{lcccccc}
\tablecaption{{\it NuSTAR} and \xmm\ spectral fitting results for the three X-ray sources}
\tablewidth{0pt}
\tablecolumns{3}
\tablehead{\colhead{Parameters}  & \colhead{N1} & \colhead{N2} & \colhead{N3}}
\startdata
\xmm\ counterpart & J183239.7$-$091610 & J183250.1$-$091401  & J183314.2$-$092109\\
$N_{\rm H} [10^{22}$\,cm$^{-2}$] & $17_{-7}^{+10}$ & $11_{-5}^{+7}$ & $6_{-4}^{+8}$ \\
$\Gamma$ & $2.1_{-0.4}^{+0.5}$ & $1.2\pm0.4$ & $0.9_{-0.3}^{+0.4}$ \\
Flux (2-20 keV)\tablenotemark{a} & $3.7_{-0.8}^{+1.5}$ & $4.1_{-0.6}^{+0.8}$ & $4.2_{-0.5}^{+0.7}$  \\
$\chi_\nu^2$ (dof) & 0.97 (34) & 0.62 (18) & 0.97 (28)
\enddata
\tablenotetext{a}{unabsorbed flux [$10^{-13}$ \eflux]. }
\label{tab:spectral_fit_3sources}
\end{deluxetable*}

\section{Discussion}
\label{sec:discussion}

\subsection{X-ray emission from gamma-ray binary candidate \hesssrc }

The X-ray spectral and timing signatures of \hesssrc\ i.e., (1) a single 
power-law spectrum up to 30~keV, (2) with a photon index $\Gamma\approx1.5$, (3) evidence for X-ray flux variation over time  and (4) a flat power
density spectrum without red noise - are consistent with the class of known TeV gamma-ray binaries \citep{Dubus2013}, suggesting that it is a 
non-accreting NS binary system. Between a neutron star and its high mass companion star, high energy emission originates 
from the shocked region where the 
stellar  and pulsar winds collide with each other \citep{Dubus2015a}. 
In this scenario, X-rays stem from synchrotron radiation from accelerated electrons, while 
inverse Compton scattering of UV photons from a massive star are responsible for gamma-rays up to the TeV band. The non-detection of a spectral break is consistent with this 
picture where the synchrotron cutoff is expected to be at much higher energy $750 (B {\rm[G]})^{-1} (d/{\rm 0.1AU})^2$~keV  for typical ranges of the magnetic field strength in the pulsar wind ($B$) 
and the binary 
separation ($d$) \citep{Dubus2013}. On the other hand, accreting NS-HMXBs normally show a spectral break at $E\sim10$-20~keV \citep{Coburn2002}. 

The 2$-$10 keV luminosity ($L_X = 2.3\times10^{33}$~erg\,s$^{-1}$),
assuming that the source is associated with the GLIMPSE9 stellar
cluster and SNR \snr\ at $\sim$4~kpc \citep{Messineo2010, Su2014}, is
similar to HESS~J0632+057 ($L_X \sim 10^{33}$~erg\,s$^{-1}$), while
other gamma-ray binaries are brighter in the X-ray band by an order of
magnitude. As \citet{Eger2016} pointed out, \hesssrc\ and
HESS~J0632$+$057 possess similar characteristics such as the faint
X-ray emission, the lack of \fermi\ GeV gamma-ray detection and the
spectral energy distribution over the X-ray and TeV gamma-ray band.
Further comparison with HESS~J0632$+$057 as well as various emission
models is not viable until an orbital period is discovered and the
high energy emission is fully characterized in different orbital
phases.  As a TeV binary, \hesssrc\ can be expected to exhibit X-ray
flares similar to other gamma-ray binaries. For example,
HESS~J0632+057 displays $\goe5\times$ X-ray flux enhancement within $<1$ month \citep{Bongiorno2011}. \swift\ monitoring of \xmmshort\ over a
year may have a good chance of detecting its orbital period.

\subsection{IR counterpart of \xmmsrc}
\label{sec:ir} 

In all TeV gamma-ray binaries, the IR and optical emission is predominantly from their massive companion stars \citep{Dubus2013}.
In the case of \hesssrc, the IR source 2MASS J18324516-0921545 with the magnitudes
$J = 15.521 \pm 0.061$, $H = 13.264 \pm 0.036$ and $K = 12.172 \pm 0.019$ coincides
with the \chandra\ position of \xmmshort\ \citep{Cutri2003, Eger2016}. 
According to the VizieR catalog, other IR surveys detected remarkably similar IR magnitudes:  UKIDSS ($J=15.359\pm0.005$, $H=13.316\pm0.002$ and  
$K = 12.118\pm0.002$) and DENIS ($J = 15.326 \pm 0.18$ and  $K = 12.080 \pm 0.16$) \citep{Lucas2008}. The weak variability of IR magnitudes such as
$\Delta J \approx 0.2$, $\Delta H \approx 0.05$ and $\Delta K \approx 0.09$
is a common signature of HXMBs \citep{Reig2015}. 
The GLIMPSE survey detected mid-IR
emission (G022.4768$-$00.1539) at magnitudes $11.393\pm0.047$ (3.6~$\mu$m), 
$11.161\pm0.068$ (4.5~$\mu$m), $11.056\pm0.090$ (5.8~$\mu$m) and $10.779\pm0.092$ (8.0~$\mu$m). 
Both the large mid-IR brightness and colors suggest that the IR source is not an AGN \citep{Stern2005, Mendez2013}. 

Given the IR magnitudes measured by 2MASS, we attempt to speculate the stellar type.
Galactic hydrogen column density ($N_{\rm H} = 1.7\times10^{22}$~cm$^{-2}$)  to \xmmshort\ 
estimated by radio surveys \citep{Willingale2013}
 leads to the optical extinction $A_V = 7.7$ using the relation
$N_{\rm H} = 2.2\times10^{21} A_V$~cm$^{-2}$ \citep{Guver2009}.
The higher hydrogen column density ($N_{\rm H} = 1\times10^{23}$~cm$^{-2}$) derived from fitting X-ray spectra indicates that this value
is a lower limit of the optical extinction since there may be additional dust absorption from a
surrounding molecular cloud.
Using the extinction ratios from \citet{Fitzpatrick2009},
we correct the IR magnitudes to $A_J=2.05$ and $A_K=0.87$. Assuming the
source distance of 4.4~kpc, we derive absolute magnitudes of $M_J=0.03$ and $M_K = -1.99$ leading to the spectral types
B8V and B1.5V, respectively  \citep{Pecaut2013}. The mismatch in the stellar types derived from the J and K magnitudes may be due to an infrared excess primarily in the K band (thus it may account for the detection of mid-IR emission) 
from warm dust or bremsstrahlung from the stellar winds. 
If the optical extinction is higher than $A_V = 7.7$ due to local dust
absorption, the IR magnitudes will be larger thus it suggests a more massive O star. However, as demonstrated for identifying hard X-ray sources discovered by \integral\ \citep{Nespoli2008, Coleiro2013}, IR spectroscopy is required to determine the exact type of a companion
star associated with \xmmshort.  

\subsection{The nature of the field sources N1, N2, N3}

\begin{figure*}[t]
\begin{center}
\epsfig{figure=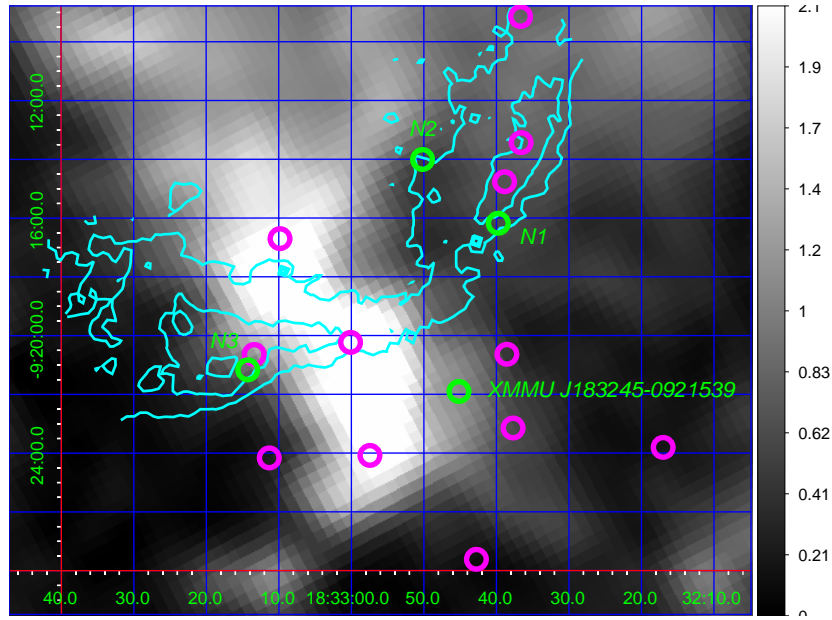,height=0.45\linewidth}
\caption{$^{13}$CO molecular line map at $v=75$~km\,s$^{-1}$ around the southern region of
the SNR \snr\ shell \citep{Su2014}. 
\xmmshort\ and the three hard X-ray sources (N1, N2 and N3) detected by \nustar\ in green circles 
are overlaid, 
while other \xmm\ sources from the 3XMM catalog \citep{Rosen2016} are indicated by magenta circles. 
Radio contours tracing the SNR \snr\ shell are shown in cyan.
}
\label{fig:13CO_map}
\end{center}
\end{figure*}

Hard X-ray detection of the three X-ray sources (N1, N2 and N3) points towards X-ray binaries 
harboring neutron star or black hole, 
magnetic CVs or pulsars as demonstrated by \nustar\ studies of Galactic point 
sources \citep{Hong2016, Fornasini2017} and serendipitous hard X-ray sources \citep{Tomsick2017}. The \xmm\ counterparts of the three hard X-ray sources are consistent with point sources and their spatial extents are 
constrained to $\la10$\asec.  
In addition, there are  about a dozen unidentified \xmm\ sources
in the region. Nearby H II regions and young stellar cluster GLIMPSE9 \citep{Messineo2010}
may account for a large number of X-ray sources.
Alternatively, some of these X-ray sources may represent 
point-like diffuse X-ray emission since they are located at the southern boundary of the \snr\ radio shell interacting  with the
molecular clouds \citep{Su2014} (see Figure~\ref{fig:13CO_map}). Fast-moving supernova fragments in a dense molecular cloud can produce compact diffuse X-ray emission features as observed in SNR IC~443 \citep{Bykov2005}. 

The three hard X-ray sources N1, N2 and N3 exhibit rather distinct \nustar\ spectra 
(Table~\ref{tab:spectral_fit}). 
They have 2--10 keV fluxes $\sim4\times10^{-13}$ \eflux\ (unabsorbed) corresponding to the
X-ray luminosity $\sim6\times10^{32}$~erg\,s$^{-1}$ assuming that these sources are associated with the SNR \snr, H~II region and GLIMPSE9 stellar cluster whose distances are 4.2--4.4~kpc 
\citep{Messineo2010, Su2014}. 
Below we discuss potential identification of 
the three \nustar\ sources largely based on the \nustar\ results. 

{\it Source N1 (XMMU~J183239.7$-$091610) --- } Among the three \nustar\ sources, N1 exhibits the softest X-ray spectra with a power-law photon index $\Gamma = 2.1_{-0.4}^{+0.5}$. 
Spectral fitting with thermal APEC model yields $kT \sim 13$~keV.  
Lack of bright IR counterparts ($K < 17$) within the \xmm\ position error circle indicates that N1 is either 
a LMXB or magnetic CV if it is a binary system. 
Another possibility is a pulsar in which case we
expect no X-ray variability over time. However,
given the large X-ray flux errors from the previous X-ray 
observations, it is unclear whether this source is variable or not.

{\it Source N2 (XMMU~J183250.1$-$091401) --- } The hard X-ray 
spectrum with $\Gamma = 1.2\pm0.4$ favors a NS-HMXB since NS-HMXBs generally have hard X-ray spectra with $\Gamma\sim1$. 
The presence of the gamma-ray binary \hesssrc\ 
and the nearby star cluster GLIMPSE9 is suggestive that some of the X-ray sources in the region may be HMXBs. 
Alternatively, N2 may be an intermediate polar since the best-fit plasma temperature is higher than $\sim30$~keV. There is an IR source with $K=14.9$ located 
within 1.3\asec\ from the \xmm\ position. Following the corrections on IR magnitudes applied to \xmmshort\ 
(\$\ref{sec:ir}), this IR source is likely a B9V star or a more massive star.  
A better X-ray source localization with \chandra\ is crucial to 
determine its IR counterpart, then follow-up IR spectroscopy can identify the companion star type.

{\it Source N3 (XMMU~J183314.2$-$092109) ---}  The \nustar\ source N3 may represent compact diffuse X-ray emission from the SNR-cloud interaction since it 
is located at the southern region of the \snr\ shell 
intersecting with molecular cloud \gmc\ and its \nustar\ spectrum fits to a hard power-law spectrum ($\Gamma\approx1$).
At the \nustar\ position of N3, there are two \xmm\ sources separated by $\sim20$\asec. It is unclear whether N3 is an extended source
overlapping the two \xmm\ sources or truly a point source. There are several \xmm\ sources overlapping with the strong $^{13}$CO line emission region. 
Such X-ray morphology is similar to that of another middle-aged SNR IC~443 ($\tau\sim3\times10^4$~years) harboring a prominent SNR-cloud interaction site with a
dozen X-ray sources \citep{Bocchino2003}. 

In IC~443, \chandra\ resolved one of the X-ray sources with a hard power-law spectrum ($\Gamma\approx1$)
to an extent of $r\sim30$\asec, and it was interpreted as a SN ejecta fragment interacting with
dense clouds \citep{Bykov2005}. Alternatively, shocked molecular clumps can emit X-rays with hard 
spectrum at a SNR-cloud interaction site such as $\gamma$ Cygni \citep{Uchiyama2002a}. However, 
this scenario is unlikely since it predicts a more extended X-ray emission ($\sim$ a few arcmin) than the 
 size ($\sim20$\asec\ or less) of the hard X-ray emission observed 
in IC~443 and N3 \citep{Bocchino2003}. 

If N3 is a SN ejecta fragment similar to that in IC~443, its X-ray luminosity ($L_X\sim10^{32}$~erg\,s$^{-1}$) indicates that the ejecta mass is likely $\goe10^{-2}M_{\odot}$ \citep{Bykov2005}. 
Given the radius ($\sim$18~pc) and age ($\sim3\times10^4$ years) of the SNR \citep{Su2014}, the estimated 
velocity ($\sim500$~km\,s$^{-1}$) of a SN fragment at the SNR shell is large enough to produce 
the observed X-ray flux similar to the diffuse X-ray features observed in IC~443 \citep{Bykov2005}.  
The angular size of such a SN fragment is expected to be $\sim$10\asec\ at the distance of \snr\, or less extended if the ejecta mass is smaller. 
Follow-up \chandra\ observation is warranted to resolve 
such small scale features. Further \xmm\ observations, improving photon statistics, may  
detect Fe emission line 
from metal-rich SN ejecta as predicted by \citet{Bykov2002}. On the other hand, if N3 is indeed a point source, its hard X-ray spectra with $kT \approx 30$~keV (APEC model) and $kT \approx 13$~keV (partially covered APEC model) suggest an intermediate polar.

\section{Summary}
\label{sec:summary}

\begin{itemize} 

\item[(1)] A 87~ksec \nustar\ observation obtained high-quality X-ray
  spectra and timing data on \xmmsrc, the likely counterpart to a new
  gamma-ray binary candidate \hesssrc. 

\item[(2)] The non-thermal X-ray spectrum of \hesssrc\ extends up to at least $\sim30$~keV with 
a power-law index of $\Gamma = 1.5$. No spectral break was observed. We found that the \nustar\ 2-10~keV flux 
is higher than that of the 2011 \xmm\ observation by a factor of $1.5_{-0.2}^{+0.4}$ (90\% c.l). 

\item[(3)] No rapid X-ray pulsation indicative of a pulsar or slow
  modulation from a binary system were detected from \xmmsrc. The flat
  power density spectrum shows no evidence for accretion.

\item[(4)] {\it NuSTAR} hard X-ray emission is detected  from three \xmm\
  sources located within the radio shell of SNR \snr. Broadband X-ray spectroscopy with \nustar\ and \xmm\ data suggests that one of these hard sources may be a supernova ejecta fragment interacting with a dense cloud, while the other two 
sources are likely X-ray binaries or magnetic CVs. A follow-up \chandra\ observation is required to identify their IR counterparts and resolve their spatial extents to smaller than 
$\sim10$\asec\ size.

\end{itemize} 

In conclusion, the X-ray spectral and timing properties of \hesssrc\
are similar to other gamma-ray binaries, especially HESS~J0632-057.
Detection of its orbital period as well as simultaneous observations
in the X-ray and gamma-ray band are the next steps to understanding
the emission mechanism and geometry.

\acknowledgements

This work was supported under NASA Contract No. NNG08FD60C, and made use of data from the \nustar\ mission, a project led by the California Institute of 
Technology, managed by the Jet Propulsion Laboratory, and funded by the National Aeronautics and Space Administration. We thank the \nustar\ Operations, 
Software and Calibration teams for support with the execution and analysis of these observations. This research has made use of the \nustar\ Data Analysis 
Software (NuSTARDAS) jointly developed by the ASI Science Data Center (ASDC, Italy) and the California Institute of Technology (USA). 

\bibliography{j1832_nustar}

\end{document}